\documentclass[acmtog]{acmart}

\usepackage{balance}
\usepackage{colortbl}
\usepackage[utf8]{inputenc}
\usepackage{booktabs} 
\usepackage{pdfpages}
\usepackage{fancyhdr,graphicx,amsmath}
\usepackage{array}
\usepackage{multirow}
\usepackage{textcomp}
\usepackage{caption}

\PassOptionsToPackage{table}{xcolor}

\setcopyright{none}

\setcitestyle{square}

\renewcommand\footnotetextcopyrightpermission[1]{} 

\usepackage[linesnumbered,ruled,vlined]{algorithm2e} 

\SetKwInput{KwInput}{Input}                
\SetKwInput{KwOutput}{Output}              

\newcommand{\sloanprt}{Sloan}
\newcommand{\mckenzie}{McKenzie}
\newcommand{\renng}{[Ng et al.]}

\begin{document}

\title{PRTT: Precomputed Radiance Transfer Textures}

\author{Sirikonda Dhawal}
\affiliation{%
  \institution{CVIT, KCIS, IIIT-Hyderabad}
  \city{Hyderabad}
  \state{Telangana}
  \country{India}
  \postcode{500032}
}
\email{dhawal.sirikonda@research.iiit.ac.in}

\author{Aakash KT}
\affiliation{%
  \institution{CVIT, KCIS, IIIT-Hyderabad}
  \city{Hyderabad}
  \state{Telangana}
  \country{India}
  \postcode{500032}
}
\email{aakash.kt@research.iiit.ac.in}

\author{P.J. Narayanan}
\affiliation{%
  \institution{CVIT, KCIS, IIIT-Hyderabad}
  \city{Hyderabad}
  \state{Telangana}
  \country{India}
  \postcode{500032}
}
\email{pjn@iiit.ac.in}

\begin{abstract}
 Precomputed Radiance Transfer (PRT) can achieve high quality renders of glossy materials at real-time framerates. PRT involves precomputing a $k$-dimensional transfer vector of Spherical Harmonic (SH) coefficients at specific points for a scene. Most prior art precomputes transfer at vertices of the mesh and interpolates color for interior points. They require finer mesh tessellations for high quality renderings. In this paper, we explore and present the use of textures for storing transfer. Using \textit{transfer textures} decouples mesh resolution from transfer storage and sampling which is useful especially for glossy renders. We further demonstrate glossy inter-reflections by precomputing additional textures. We thoroughly discuss practical aspects of transfer textures and analyze their performance in real-time rendering applications. We show equivalent or higher render quality and FPS and demonstrate results on several challenging scenes.
\end{abstract}

\ccsdesc[500]{Computing methodologies~Rasterization}
\ccsdesc[500]{Computing methodologies~Ray tracing}
\ccsdesc[300]{Computing methodologies~Visibility}

\keywords{Transfer, fragment-shader, Visibility, Textures}

\begin{teaserfigure}
  \includegraphics[width=\linewidth]{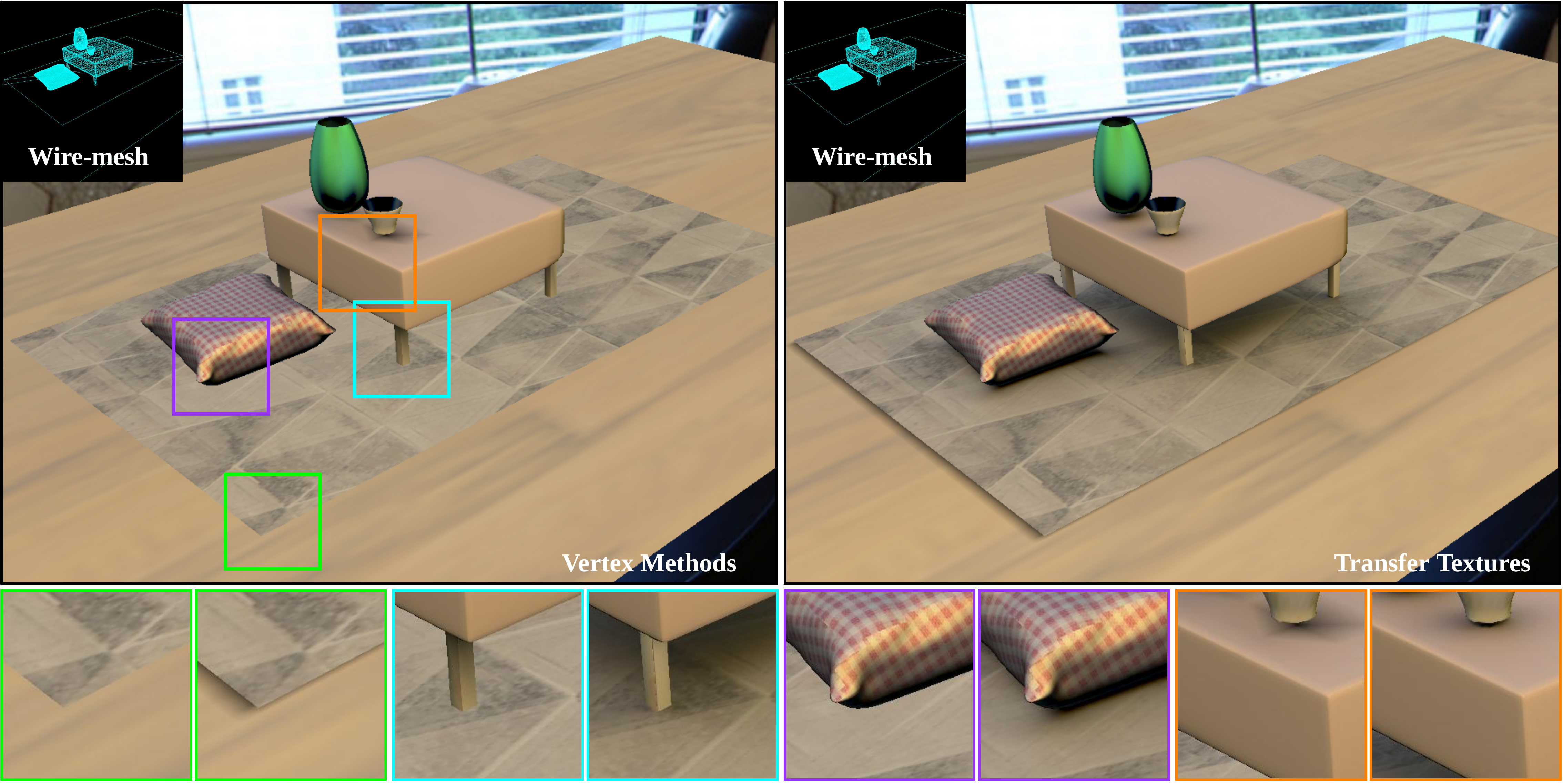}
  \caption{We present transfer textures to decouple mesh resolution from transfer storage and sampling. Our method precomputes and stores transfer in a texture that is sampled in a fragment shader. The above scene has a minimally tessellated rug, floor and table (shown in wireframe insets). Traditional triple product method on vertex shader (left) is unable capture shadows due to insufficient sampling of the transfer function. In contrast transfer textures run on the fragment shader and accurately captures all details for the same tessellation. Note that the scene has spatially varying color and roughness.}
  \label{fig:teaser}
\end{teaserfigure}

\fancyfoot{}
\maketitle
\thispagestyle{empty}

\section{Introduction}
Ray Tracing is the method of choice for high-fidelity image generation. However, it is computationally expensive for real-time applications where only a few milliseconds of rendering time is available per-frame. Precomputed Radiance Transfer (PRT) offloads expensive computations of ray tracing to a pre-computation step, after which the stored data can be utilized for real-time photorealistic rendering. PRT uses Spherical Harmonic (SH) lighting \cite{prt_ravir} to efficiently store and render complex effects in both video games and offline rendering for movie production \cite{prt_movie}. Recently, PRT has been extended to efficiently handle area light sources \cite{belcour_sh_area_light,analytic_sh_poly_light,analytic_sh_many_poly_lights} further increasing its utility.

The PRT framework proposed by \cite{prt_sloan} pre-computes the \textit{transfer} function and stores it in SH basis at vertices of the scene, possibly with compression \cite{pca_prt_sloan}. For a band $l$ SH projection with $k=l^2$ coefficients, this amounts to storing a $k$-dimensional vector for diffuse and $k \times k$-dimensional matrix for glossy materials at each vertex. Using the Triple Product formulation \cite{triple_ren}, it is possible to still store a $k$-dimensional vector per-vertex for glossy materials by additionally storing a global three-dimensional \textit{tripling coefficient matrix}. A further modification of triple products fixes the light \cite{sloan2008stupid} and requires the storage of a two-dimensional global matrix. The advantage of this method is that it retains a compute complexity of $O(k^2)$ as opposed to $O(k^{5/2})$ of traditional triple products. Recently, the work of \cite{fast_sh_lingqi} reduces this time complexity to $O(k^{3/2})$, although a real benefit is gained only for large $k$. 

The above PRT approaches store the transfer vectors/matrices on vertices of the mesh. While rendering, color is evaluated at each vertex and interpolated for internal points. This necessitates a reasonably dense mesh tessellation (high mesh resolution) for high-quality renders. 
There exists few methods like textured hierarchical PRT \cite{mckenzie2010textured} and PRT of D3D9 framework \cite{d3d9} which leverage the continuous texture space to store transfer. The work of \cite{iwanicki} also uses texture space similar to \cite{mckenzie2010textured} with a specific focus on shadows. These methods focus only on diffuse reflection and do not demonstrate glossy reflection and inter-reflections with textures. Directly extending above mentioned methods for glossy renders requires $k^2$ coefficients per each texel resulting in heavy texture storage, because they use the original PRT formulation of \cite{prt_sloan}. These extensions are thus infeasible in real-time scenarios (Table \ref{tab:scene_config}). Can methods like Textured-PRT be extended to handle glossy materials, inter-reflections, etc., at real-time rendering rates?

In this paper, we present {\em precomputed radiance transfer textures} (transfer textures) to efficiently store the transfer values using $k$ coefficients per texel for both diffuse and glossy reflections. Our method uses the triple product formulation of PRT \cite{triple_ren}. Transfer textures store more finely sampled transfer values and can evaluate color for each fragment in a fragment shader. This improves render quality even for coarsely tessellated meshes (Fig. \ref{fig:teaser}). They can also support high-quality local effects like glossy renders, inter-reflections, and normal maps. Higher texture resolution results in greater precomputation effort but higher render quality at fixed render times. This provides a way to balance computation effort and rendering quality and to trade one off for the other. Texture space techniques like mip-mapping and texture sets can be used for greater efficiency. We describe methods to correctly and efficiently compute transfer textures and show real-time framerates and superior render quality at low mesh tessellations for several scenes. We further formulate and demonstrate inter-reflections using transfer textures by completely fixing the light. We also show transfer texture facilitate local effects like normal mapping. Finally, we compare and analyze run-times and storage against vertex-based approaches with the triple product method.

\section{Related Work}

\paragraph{Precomputed Radiance Transfer (PRT)} PRT was first proposed by \cite{prt_sloan}, where they projected environment lighting and light transport to SH basis for dynamic lighting for diffuse and glossy materials. Since then, PRT and SH have received lot of attention to efficiently compute SH basis \cite{snyder2006code}, efficient rotation of SH \cite{zh_to_sh_rotation}, compressing SH basis \cite{pca_prt_sloan}, microfacet BRDFs \cite{microfacet_brdf_prt} and extending PRT for dynamic scenes \cite{dynamic_prt}. More recently, \cite{analytic_sh_poly_light} extend PRT to support area lights achieving real-time frame rates for a few light sources. This work was later extended to support a large number of area lights while maintaining real-time frame rates \cite{analytic_sh_many_poly_lights}. All of these methods were either orthogonal to the core PRT framework or extended the core framework to support additional scenarios. In contrast, we improve the core PRT framework to compute and store transfer on a texture instead of at vertices.

\paragraph{Triple Products.} Triple products naturally arise in computer graphics in the rendering equation. Triple products in wavelet basis and spherical harmonics have been studied in depth \cite{triple_ren}. Today, state-of-the art in PRT uses triple products for dynamic relighting for diffuse and glossy scenes. By itself, triple product method has a compute complexity of $O(k^3)$. This method can be made more computationally efficient by fixing the lighting \cite{sloan2008stupid}. Specifically, triple products with fixed lighting in SH based PRT achieves a compute complexity of $O(k^2)$ and per-vertex storage of $k$-dimensional vector. We augment the triple product method with transfer textures and demonstrate superior rendering quality and real-time framerates.


\paragraph{Transfer stored on Textures.}
Storage of transfer on a texture was first suggested by \cite{prt_sloan}. \cite{mckenzie2010textured,iwanicki} formally demonstrated transfer storage on textures. These methods mainly focused on diffuse materials without inter-reflections. In this work, we focus primarily on glossy materials and also demonstrate inter-reflections with transfer textures.

\section{Background on PRT with Triple Products}
\label{sect:prt_intro}
To make our document self-contained, we first review and discuss Precomputed Radiance Transfer with triple products, which is the current state of the art in PRT. The inspiration for the choice of using triple product formulation is discussed in Sect. \ref{subsec:memory_req}. The rendering equation for direct lighting at point $p$ is given by:
\begin{equation}
    \label{eq:rendering}
    B^p(\omega_o) = \int_{\Omega} L(\omega_i) \rho^p(\omega_o, \omega_i) V^p(\omega_i) (\omega_i \odot n) d\omega_i, 
\end{equation}
where $\omega_o$ is the direction towards the viewer from $p$, $\omega_i$ is the incoming direction on the unit hemisphere $\Omega$ and $n$ is the surface normal at $p$. $B^p$ is the reflected radiance in direction $\omega_o$, $L$ is the incoming environment light from $\omega_i$, $V^p$ is the binary visibility function and $\rho^p$ is the Bi-directional Reflectance Distribution Function (BRDF). Eq. \ref{eq:rendering} is decomposed into the lighting $L$ and transfer $T^p(\omega_i) = V^p(\omega_i) (\omega_i \odot n)$ which are then projected to the SH basis with coefficients $T^p_i$ and $L_i$ respectively. Triple products \cite{triple_ren} formulate transferred radiance as:
{\small \begin{equation}
    \label{eq:trip_prod}
    L^{p}_k = \int_{S^2} y_k(\omega) \left ( \sum_{i=1}^{n^2} T^p_i y_i(\omega) \right ) \left ( \sum_{j=1}^{n^2} L_j y_j(\omega) \right ) d\omega = \sum_{ij} \tau_{ijk}T^p_{i}L_{j},
\end{equation}}
 $B^p$ where $\tau_{ijk}$ is the triple product tensor (tripling coefficient matrix). The final color is calculated by convolution of $L_k^{p}$ with BRDF coefficients and evaluation at reflection direction.
The above formulation for PRT results in a $k$-vector for transfer stored at each point and $O(k^{5/2})$ compute \cite{triple_ren} required for rendering. The compute efficiency can be further improved to $O(k^2)$ by fixing the light and precomputing a global product matrix: $(M)_{ik} = \sum_{j} \tau_{ijk} L_{j}$ \cite{sloan2008stupid}.

\section{Transfer Textures}
\label{sect:main_transfer_tex}
In this section, we begin with a description of computing and storing a band $l$ SH projection of transfer on a texture (Sec \ref{sect:compute_transfer_tex}). Next, we show how inter-reflections can be pre-computed and incorporated in our framework (Sec \ref{sect:inter_reflections}). Our implementation is described in Sec \ref{sect:impl}. Our approach achieves real-time frame rates and better render quality, especially on low tessellation meshes, as shown in Sec \ref{sect:results}.
\begin{algorithm}[htb]
    \caption{Pre-computing and storing the transfer texture.}
    \label{alg:transfer_texture}
\DontPrintSemicolon
  \KwInput{$\mathcal{M}, w, h, l$: Mesh $\mathcal{M}$, width $w$ \& height $h$, SH band $l$.}
  \KwOutput{$T_o$: Precomputed transfer texture}
  $T_o$ $\leftarrow$ Texture($w$, $h$, $l$) \tcp*{Init. texture.}
  $\mathcal{G}$ = OpenGL($\mathcal{M}$) \tcp*{G-Buffer}
  
  \For{t \textbf{in} $T_o$}
  {
    point = $\mathcal{G}$[t.x][t.y].vertex \\
    normal = $\mathcal{G}$[t.x][t.y].normal \\
    $V$ = ComputeTransfer(point, normal) \tcp*{Path tracing}
    $V_{sh}$ = SHProject($V$) \\
    $T_o$[t.x][t.y] = $V_{sh}$ \\
  }
  $T_o$ = Dilate($T_o$, 3)
\end{algorithm}

\subsection{Pre-computing Transfer Textures}
\label{sect:compute_transfer_tex}

The computation of transfer involves shooting multiple rays from a point $p$ in the scene and then evaluating and projecting the transfer to the SH basis. For transfer textures, there are $N$ scene points $p$ corresponding to each pixel $t$ in the texture. The mapping between $t$ and $p$ is defined by the UV coordinates.
To efficiently compute the transfer texture, we leverage G-Buffers(Alg. \ref{alg:g_buffer}) to interpolate vertex positions and normals based on their corresponding UV-Coordinates (Alg. \ref{alg:transfer_texture}, line 2). Next, we read the G-buffer and the scene geometry and evaluate the transfer function for each pixel in the buffer (Alg. \ref{alg:transfer_texture}, lines 4-6). The transfer obtained is then projected to SH basis and stored at the same pixel location in a initially empty texture $T_o$ (Alg. \ref{alg:transfer_texture}, lines 7-8). Finally, $T_0$ is dilated to ensure that all points inside a triangle receive a transfer value. At run-time, we fetch transfer $T_o$ and use it with the triple product formulation to obtain $B^p$.

\begin{algorithm}[ht]
    \caption{G-Buffer pass.(Vertex and fragment code)}
    \label{alg:g_buffer}
\DontPrintSemicolon
    \SetKwProg{Fn}{Program}{:}{}
    \Fn{VertexShader} {
        vec4 gl\_Position; \tcp*{In-built variable}
        \textbf{in} vec3 p, n; \tcp*{Scene Point, Normal}
        \textbf{in} vec2 uv; \tcp*{UV co-ordinates}
        \textbf{out} vec3 vertex; \tcp*{Interpolated in Frag.}
        \textbf{out} vec3 normal; \tcp*{Interpolated in Frag.}
        \SetKwProg{Fn}{void}{:}{}
        \Fn{main()} {
            gl\_Position = vec4(uv.x, uv.y, 0.0, 1.0); \\
            vertex = p; normal = n; \\
        }
    }
    \SetKwProg{Fn}{Program}{:}{}
    \Fn{FragmentShader} {
        \textbf{in} vec3 vertex;\\
        \textbf{in} vec3 normal; \\
        \textbf{out} vec4 gPos; \tcp*{G-Buffer}
        \textbf{out} vec4 gNorm; \tcp*{G-Buffer}
        \SetKwProg{Fn}{void}{:}{}
        \Fn{main()} {
            gPos = vec4(vertex,1.0); \\
            gNorm = vec4(normal, 1.0); 
            \tcc{\emph{w} $\rightarrow$ alpha channel}
        }
    }
\end{algorithm}

\subsection{Pre-computing transfer textures for inter-reflections}
\label{sect:inter_reflections}

\begin{figure}[htb]
    \centering
    \includegraphics[width=\linewidth]{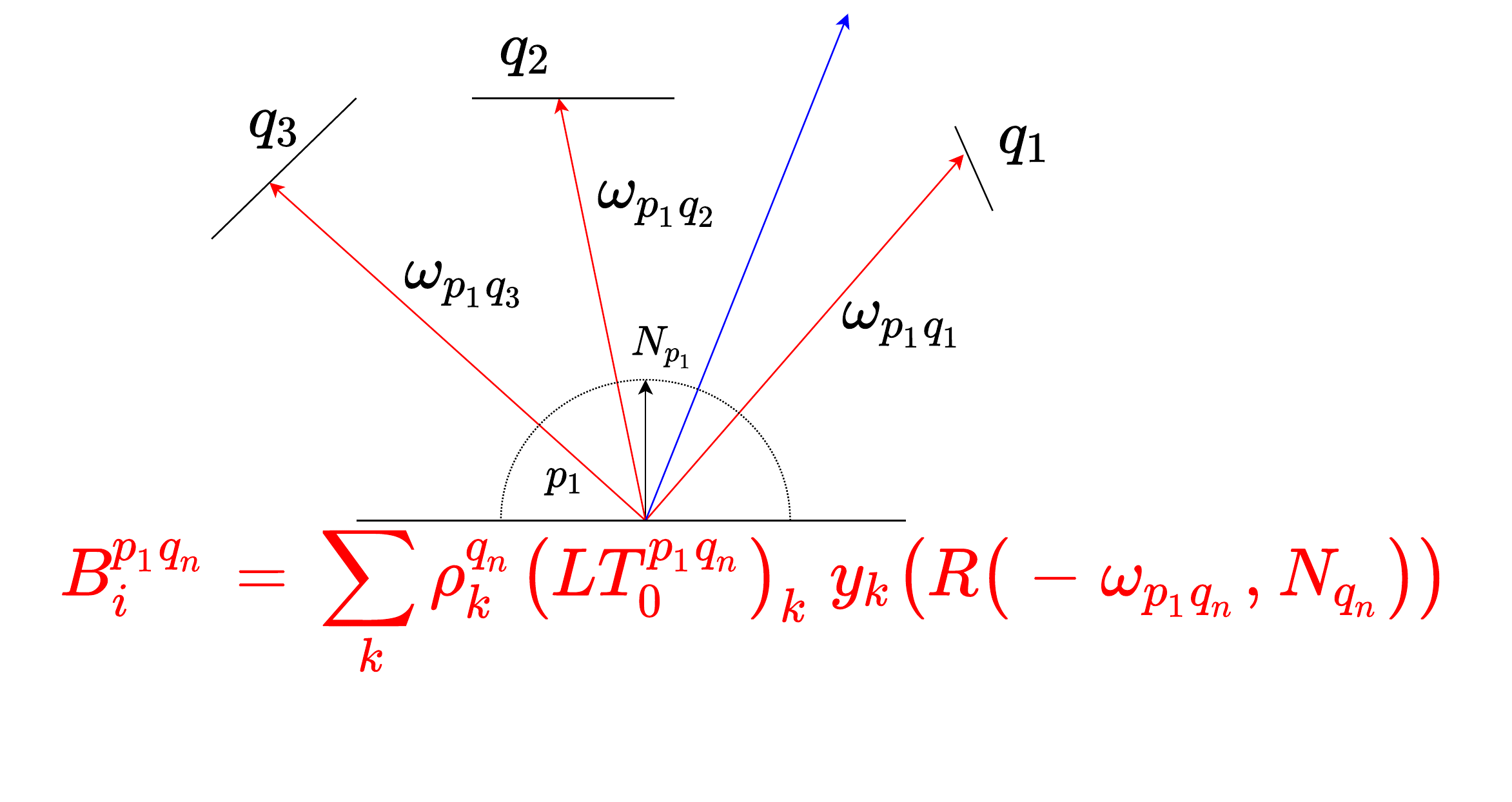}
    \caption{\textbf{Inter-reflections:} The radiance $B^{p_1q_n}$ (red lines) towards a point $p_1$ from a secondary hit-point $q_n$ can be computed by first fetching transfer at $q_n$ using the zero-bounce transfer texture $T_o$, applying the light $L$ followed by convolution with the BRDF at $q_n$ and evaluation at reflected direction along the normal at $q_n$. $B^{p_1q_n}$ forms an indirect environment map which is projected to SH and stored at $p_1$ in an additional texture.}
    \label{fig:interreflection_method}
\end{figure}

Inter-reflected radiance $B_i^p$ at point $p$ can be modeled as:
\begin{equation}
    \label{eq:inter_rendering}
    B_i^p(\omega_o) = \int_{\Omega} (1-V^p(\omega_i)) B^{pq}(x, \omega_i) \rho^p(\omega_o, \omega_i) (\omega_i \odot n) d\omega_i,
\end{equation}
where $B^{pq}$ is the radiance from a secondary hit-point $q$ towards $p$ and $B_{i}^{p}$ is the inter-reflected radiance ~\cite{prt_sloan}. First, we factor out $1-V^p(x, \omega_i)$ by only integrating over rays which hit some geometry. For a scene point $p_1$ and a secondary hit $q_1$, the radiance $B^{p_1q_1}$ can easily be precomputed given a zero-bounce transfer texture $T_o$ from Alg. \ref{alg:transfer_texture} (See Fig. \ref{fig:interreflection_method}).
The radiance from $q_1$ towards $p_1$ is obtained using the triple product formulation by fetching $T_o$ to obtain transfer at $q_1$.

This is done for all hit-points from $p_1$. This radiance now forms an \textit{indirect environment map} for the point $p_1$, which is then projected to SH  basis resulting in a $k$-vector $B^{pq}_i$, which is stored in a separate \textit{one-bounce inter-reflection} texture $T_1$.
At run-time, the inter-reflected radiance is obtained by convolving $B^{pq}_i$ fetched from $T_1$ with the BRDF SH $\rho^p_i$ and evaluating at the reflection direction. The final color is given as: $B^p(\omega_o) + B_i^p(\omega_o)$. Alg. \ref{alg:transfer_texture} can be easily extended to compute the second bounce texture $T_2$ and so on. The number of textures required is linear in the number of bounces in this setting, and the final color is just their summation.

\subsection{Handling dense UV-packing}
\label{sect:texture_sets}
In the previous section we described methods for efficient computation of transfer textures. Usage of these textures requires UV co-ordinates each vertex to be defined. To obtain UV unwrapping of scene geometry, we used \emph{Smart UV-Unwrap} or \emph{Light Map Pack} from Blender 3D \cite{Blender}. One caveat with UV unwrapping is that dense packing of UV islands may cause overlaps which manifest as rendering artefacts. \emph{Smart UV-Unwrap} does not guarantee non-overlapping islands while \emph{Light Map Pack} leads to texture wastage and tiny pixel coverage for some parts of the geometry. In such scenarios, texture-sets are beneficial. 
\begin{figure}[htb]
    \centering
    \includegraphics[width=\linewidth]{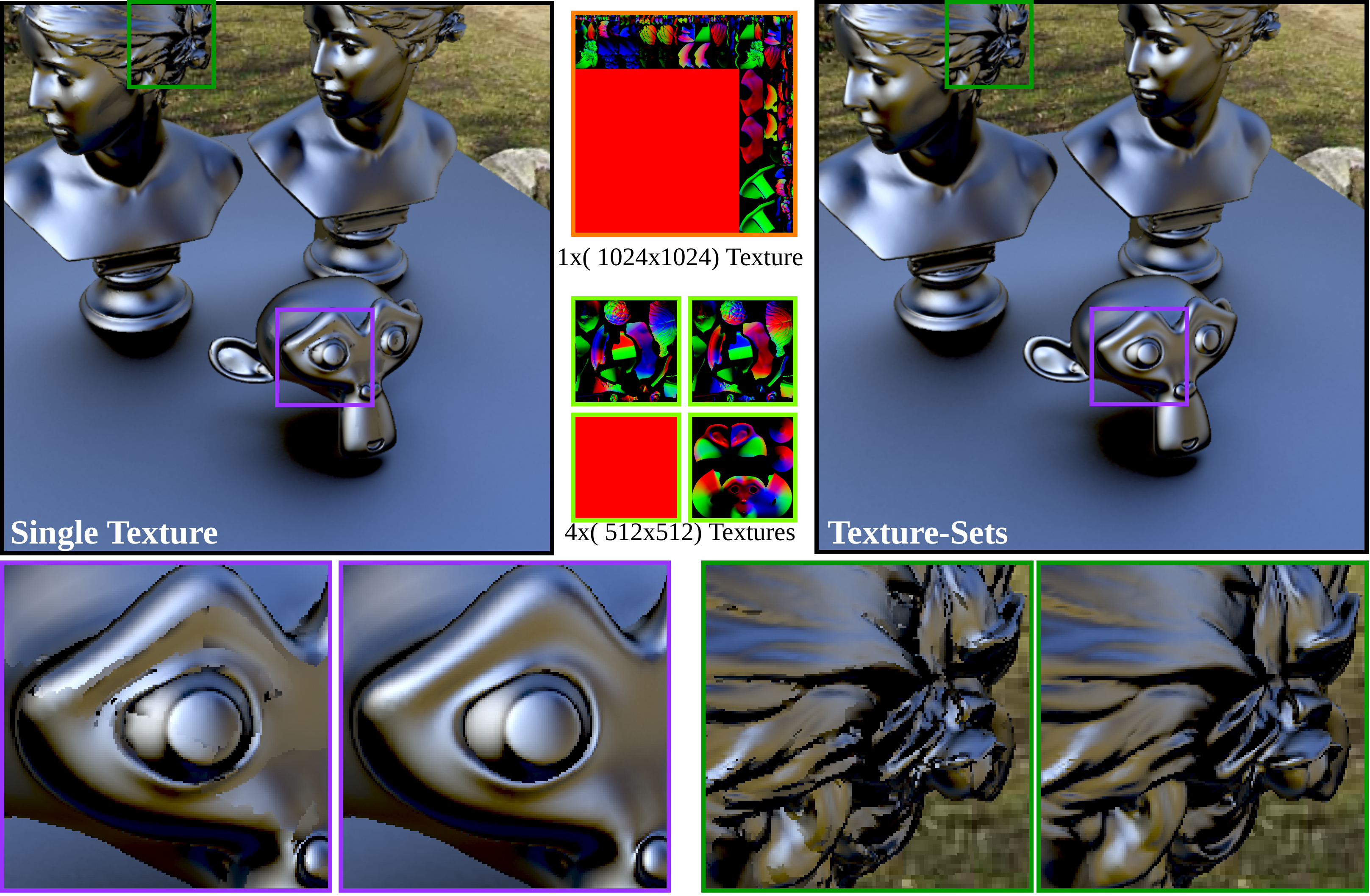}
    \caption{\textbf{Texture sets:} TRM:Two Roza, one Monkey, We demonstrate the use of texture-sets with TRM(right) where 4 small textures are assigned to each piece of geometry against the use of single texture in case of TRM(left). The artefacts can be seen clearly in the Monkey's eyes and Roza's hair as depicted in insets. Note that in both cases, the memory requirements are the same (single $1024\times1024$ texture v/s four $512\times512$ textures).}
    \label{fig:texture_sets}
\end{figure}
Consider an example scene as shown in Fig. \ref{fig:texture_sets}. This scene contains 441K triangles, all of which are packed into a single $1024\times1024$ texture (Fig. \ref{fig:texture_sets}, left). As shown in the insets, this leads to artefacts. A better approach is to use texture-sets, which means assigning individual textures to each object in the scene (Fig. \ref{fig:texture_sets}, right). In this case, each UV island can occupy the entire space of the texture thus eliminating artefacts.

\section{Implementation Details}
\label{sect:impl}
We implement Alg. \ref{alg:g_buffer} in Python using the ModernGL \cite{moderngl} framework. We generate and store the resulting G-buffers for each scene in a pre-process step. Alg. \ref{alg:transfer_texture} is implemented in Python and uses Embree \cite{embree} for efficient ray intersection tests. We project to band $l=5$ (25 coefficients) real spherical harmonics. As mentioned in Sect. \ref{sect:compute_transfer_tex} dilation is required to ensure that all points in the scene receive a transfer value. Experimentally, we found a dilation of three to be sufficient which may need adjustment depending on the scene complexity. The time taken for generating transfer textures for a scene like in Fig. \ref{fig:teaser} is approximately three hours.

Our real-time renderer is also implemented in the ModernGL framework. We implement the triple product (TP) and triple product with fixed light (TPFL) methods augmented with our transfer textures. Rendering is done in the fragment shader using the generated transfer textures for the respective scene. We render all scenes with glossy materials with spatially varying roughness on a workstation with an NVIDIA RTX 3090 with a resolution of 1920$\times$1080. An important detail is that we use the early depth pass to prune fragments that are not visible thus avoiding unnecessary computations.
We use a texture resolution of 1024$\times$1024 texture as we have found it to be best trade-off in between memory and quality for our scenes.

\begin{table}[H]
  \begin{center}
    \caption{Scene configurations. We list all scenes used in this paper, with their corresponding number of triangles and FPS with triple product (TP) and triple product fixed light (TPFL) methods on both vertex and fragment shaders (with transfer textures). Our approach achieves real-time framerates on all scenes.}
    \label{tab:scene_config}
    \begin{tabular}{l|c|c c|c c} 
      \textbf{Scene} & \textbf{\# tris.} & \multicolumn{2}{c|}{\textbf{Vert. (Trad.)}} & \multicolumn{2}{c}{\textbf{Frag. (Ours)}}\\
      & & TP & TPFL & TP & TPFL\\
      \hline
      Dragon (Fig \ref{fig:results}) & 1.3M & 3.62 & 41.2 & 5.2 & 151.2 \\
      TRM (Fig \ref{fig:results}) & 441K & 10.2 & 116.2 & 15.2 & 202.9\\
      Room (Fig \ref{fig:teaser}) & 21K & 352.3 & 2432.7  & 83.6 & 568.2\\
      Plants (Fig \ref{fig:results}) & 18K & 363.2 & 2597.6 & 6.7 & 168.3 \\
    \end{tabular}
  \end{center}
\end{table}
\section{Results \& Evaluation}
\label{sect:results}
\begin{figure*}
    \centering
    \includegraphics[scale=0.25]{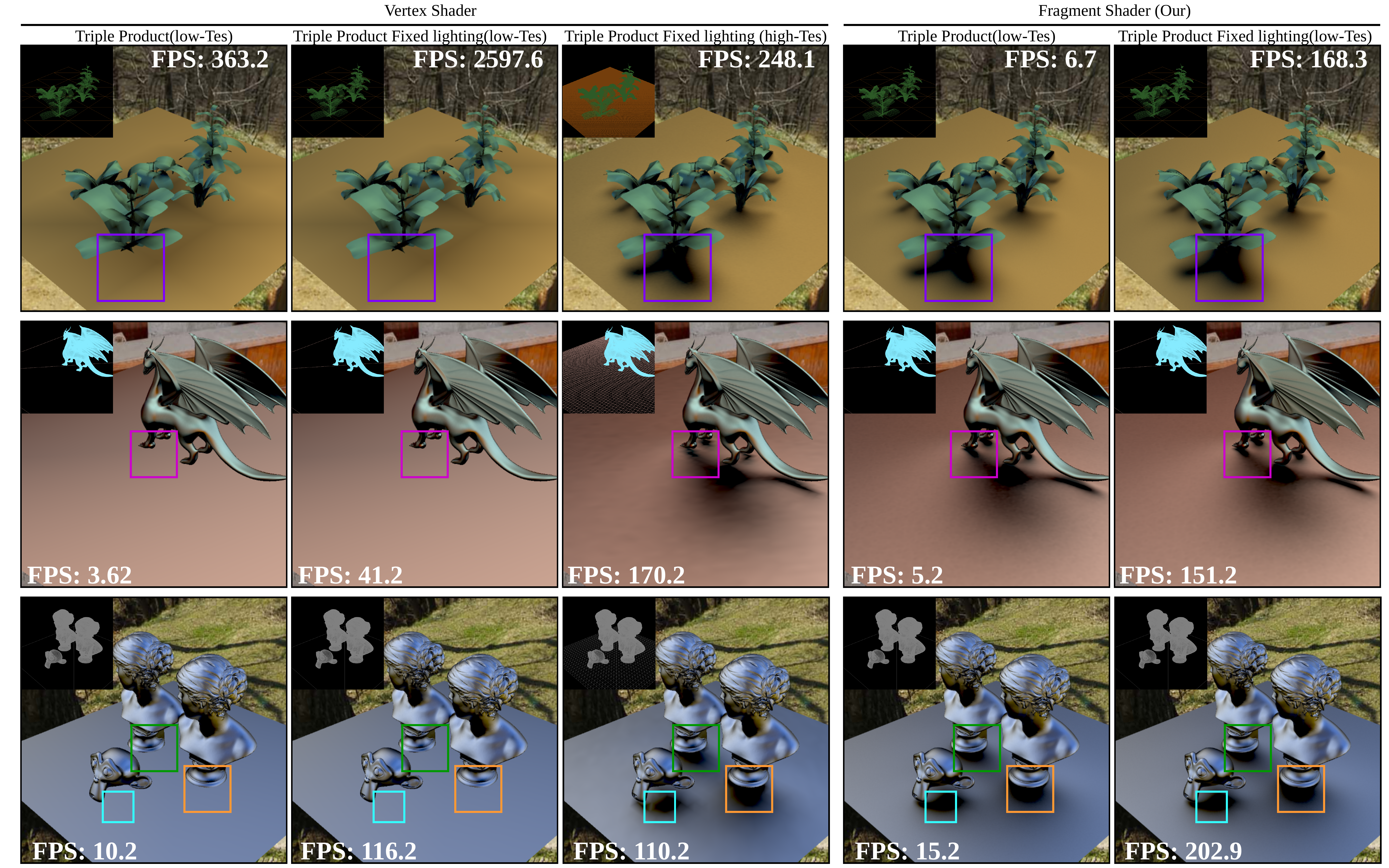}
    \caption{We show results of TP and TPFL on the fragment shader using our transfer textures (Bottom). We compare renderings with traditional vertex shader based approaches on the top. For minimal tessellations, our method accurately renders shadows whereas previous methods are unable due to insufficient sampling of transfer. The third row (high-tessellation) shows that renderings using traditional methods approximately approach the quality of transfer textures on addition of more vertices. Note that low-FPS in case of {Dragon low-tessellation} and {TRM low-tessellation} in vertex based TP and TPFL is due to their high resolution geometry.}
    \label{fig:results}
\end{figure*}

In this section, we present glossy rendering results including inter-reflections using transfer textures on the fragment shader. We compare the renderings with traditional vertex shader based approaches. We also discuss and demonstrate the use of normal maps with transfer textures which is not possible with traditional vertex based PRT. Finally, we analyze the memory requirements and give a lower bound of FPS for tranfer texture usage in a fragment shader. Rendering results are demonstrated on four scenes whose statistics and performance comparisons are given in Table \ref{tab:scene_config}. 

\subsection{Glossy rendering \& Inter-reflections}
\begin{figure*}
    \centering
    \includegraphics[width=0.9\linewidth]{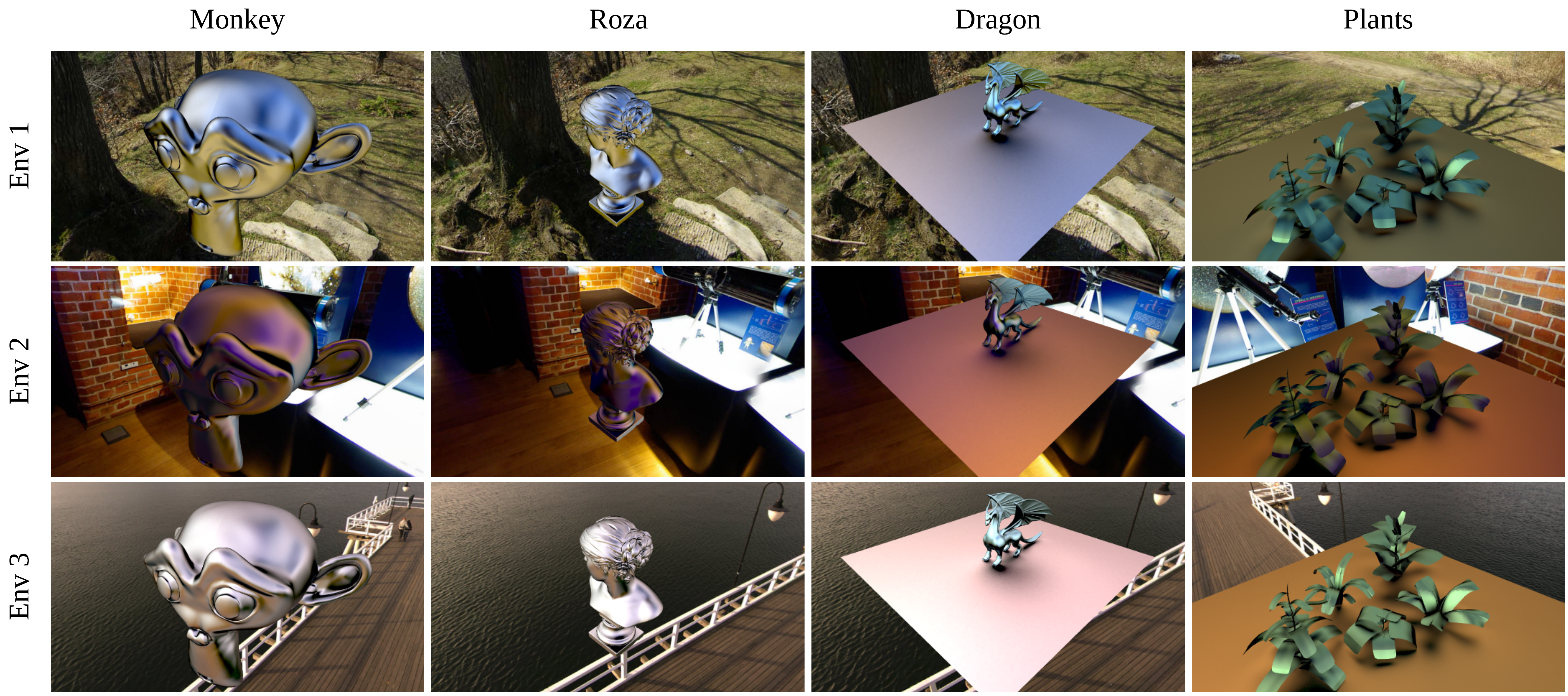}
    \caption{Results of our transfer textures method on different light settings}
    \label{fig:results_1}
\end{figure*}

\begin{figure*}
    \centering
    \includegraphics[width=0.9\linewidth]{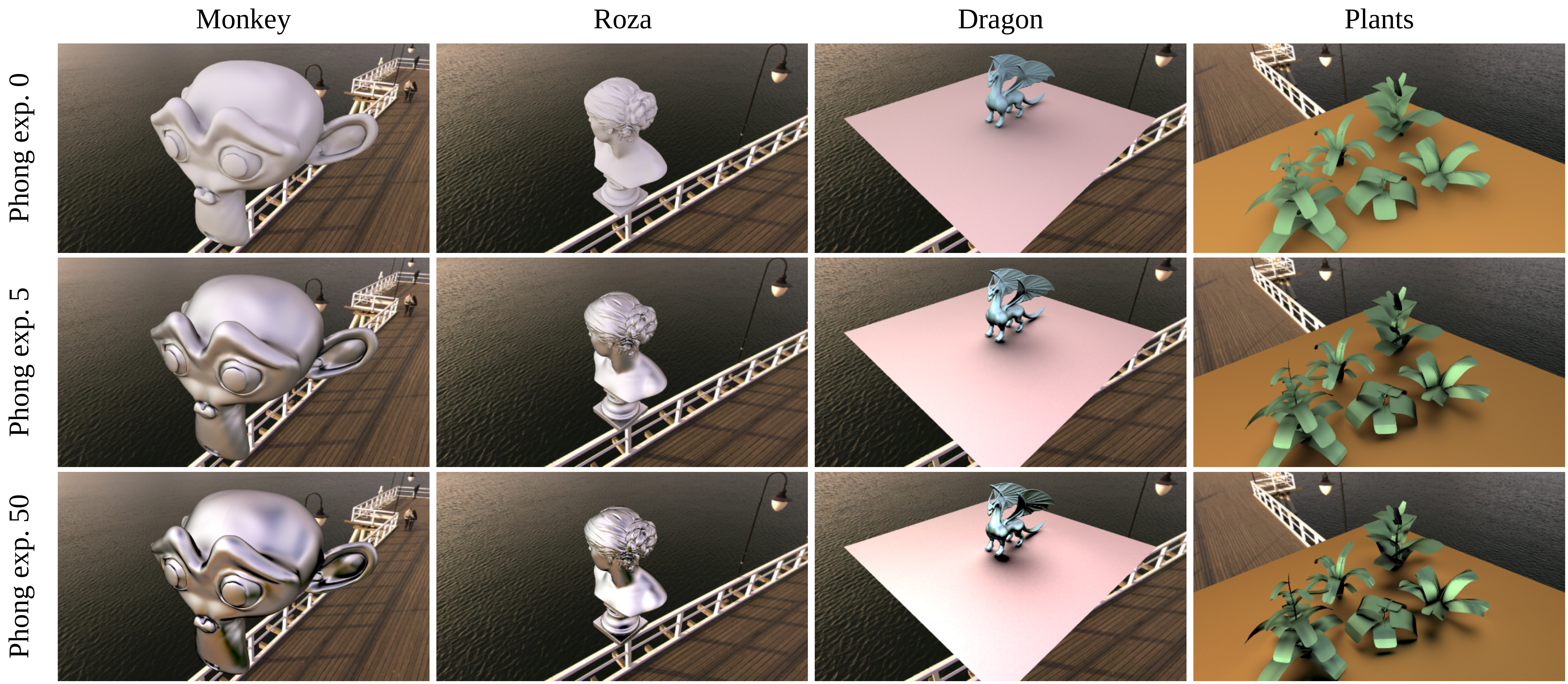}
    \caption{Results of our transfer textures method with different Phong BRDFs with same light}
    \label{fig:results_2}
\end{figure*}

Fig. \ref{fig:results} shows the renders for three scenes: \textit{Plants}, \textit{Dragon} and \textit{TRM (two Roza, one Monkey)}. All scenes have a ground plane, which is minimally tessellated, as shown in the wireframe insets. The TP and TPFL methods on vertex shader are unable to capture proper shadows on the ground plane due to sparse sampling of the transfer function. In contrast, the TP method on the fragment shader using our transfer textures properly reproduces shadows on the plane, albeit at a very low FPS. The TPFL method with transfer textures also achieves a similar render quality at a higher FPS. We note that the TP/TPFL methods on vertex shader approach the render quality of our transfer textures with a highly tessellated ground plane, as shown in the \textit{high-tessellation} renderings. We note that this requires the addition of \emph{redundant} vertices. We further note that such situations frequently arise in production, for example with walls in a room or any large surface with minimal curvature (Fig \ref{fig:teaser}). In such cases, all previous PRT methods on vertex shaders require the addition of avoidable vertices to store the transfer on leading to drop in performance, as opposed to our transfer textures method. Additional renders with different phong exponents and environments maps for four different scenes are shown in Fig. \ref{fig:results_1} \& \ref{fig:results_2}.
\begin{figure}
    \centering
    \includegraphics[width=\linewidth]{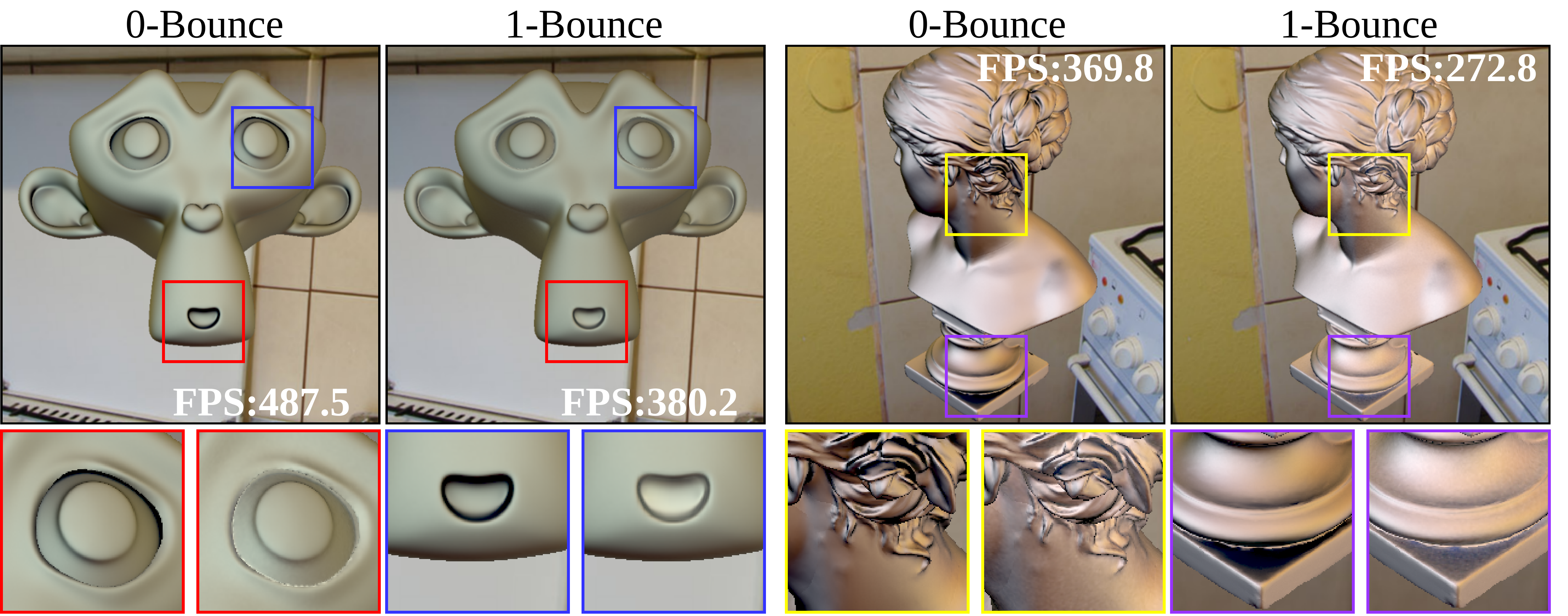}
    \caption{We demonstrate inter-reflections with transfer textures on two scenes: Diffuse Monkey (left) and glossy Roza (right). Renders with inter-reflection maintain real-time frame-rates, albeit slightly lesser than zero-bounce renderings.}
    \label{fig:interreflections}
\end{figure}
Next, we demonstrate inter-reflections using transfer textures with the method described in Sec \ref{sect:inter_reflections}. The zero-bounce and one-bounce renders with their corresponding FPS are shown in Fig. \ref{fig:interreflections} for two scenes: \textit{Monkey} and \textit{Roza}. Because of extra texture fetch, convolution and evaluation operations the FPS with inter-reflections is slightly lower, albeit still real-time. As described in Sect. \ref{sect:inter_reflections}, additional bounces can be added with additional pre-computed textures. 

\begin{figure}[t]
    \centering
    \includegraphics[width=\linewidth]{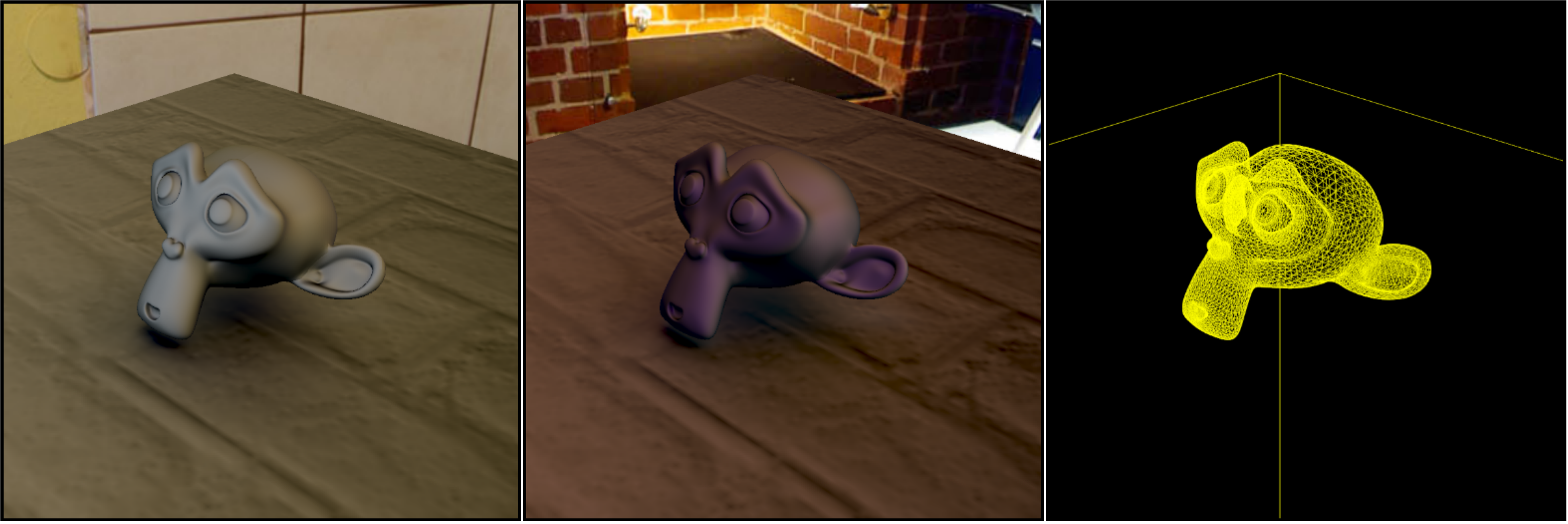}
    \caption{\textbf{Normal Maps:} Transfer textures make it possible to use the shading normals from normal maps instead of the geometric normals. This is difficult to achieve with traditinal vertex based PRT. The above scene shows a minimally tessellated ground plane (wireframe, right) with a normal map. The scene is rendered with TPFL and transfer textures. All normal map details are preserved.}
    
    \label{fig:normal_map}
\end{figure}
\label{subsec:normal_maps}

\subsection{Normal Maps} 
Transfer textures make it possible to use normals maps during precomputation. This translates to lesser vertices during rendering as finer detail can instead be embedded in the normal map. Consider precomputation in traditional vertex based PRT. In this case if a normal map is applied, it only ever affects the transfer at those vertices thus loosing detail within each face when using high frequency normal-texture. With transfer textures we can output the \textit{shading normal} from the normal map instead of the \textit{geometric normal} in the G-buffer during precomputation. In Alg. \ref{alg:transfer_texture} line 6, the transfer will then be computed at the shading normal instead. Since this texture is used to fetch transfer during rendering, all normal map details are preserved. We show renderings with normal maps in Fig. \ref{fig:normal_map}. The detail on the floor is due to the normal map without any additional vertices, as can be seen in the wireframe insets.

\subsection{Memory Requirements}
\label{subsec:memory_req}
\mckenzie \ demonstrated textures with diffuse PRT using \sloanprt's formulation. Consider directly implementing \sloanprt's glossy formulation with textures instead. This amounts to storing a $k \times k$ matrix per texel which quickly becomes intractable, even for reasonably small textures. Thus augmenting the triple product formulation to transfer textures is a clear choice. The memory requirements for vertex as well as texture (fragment) based approaches is shown in Table \ref{tab:memory_requirements}. The former's memory requirements depend on the scene complexity whereas it is constant for textures. Furthermore, a direct extension of \sloanprt's method to textures is infeasible, as shown in the fifth column (2.5 GB per texture).

\begin{table}[H]
  \begin{center}
    \caption{Memory requirements for vertex based and transfer texture based approaches for a $1024 \times 1024$ texture. Note that \mckenzie \ uses \sloanprt's approach which results in large textures for glossy rendering.}
    \label{tab:memory_requirements}
    \begin{tabular}{l|c|c c|c c} 
    \textbf{Scene} & \textbf{\# tris.} & \multicolumn{2}{c|}{\textbf{Vert. Mem.}} & \multicolumn{2}{c}{\textbf{Tex. Mem.}}\\
      & & \sloanprt & \renng & \mckenzie & Ours \\
      \hline
      Room & 21K & 64MB & 2.5MB & 2.5GB & 100MB\\
      Dragon & 1.3M & 5.2GB & 215.8MB & 2.5GB & 100MB \\
      TRM & 441K & 2.3GB & 139.3MB & 2.5GB & 100MB\\
      Plants & 18K & 64MB & 2.5MB & 2.5GB & 100MB \\
    \end{tabular}
  \end{center}
\end{table}

\subsection{Lower Bound on FPS} Since transfer textures are used in fragment shaders with an early depth pass, we achieve a lower bound on the FPS. The computation is roughly the same for each fragment and the worst case is when all fragments contain some geometry to be processed and rendered. This is in contrast to vertex based approaches, where run-time depends on the number of vertices in the scene. We demonstrate this in Fig. \ref{fig:results} in the \textit{TRM} (441K verts) and \textit{Dragon} (1.3M verts) scenes. Here, the FPS is lower for vertex based approach as compared to fragment based approach with transfer texture, in both TP and TPFL. 

\section{Conclusions, Limitations \& Future work}

In this paper, we presented \textit{precomputed radiance transfer textures} for decoupling mesh tessellation from transfer sampling and storage for glossy rendering. We described methods to efficiently and correctly compute these textures and also demonstrated incorporation of inter-reflections using additional precomputed textures. We compared our renderings with traditional vertex based PRT approaches and thoroughly analyzed the memory requirements of transfer textures. We demonstrated real-time framerates for rendering with transfer textures on the fragment shader and superior render quality for minimally tessellated meshes. Additionally, we gave a lower bound on the FPS which will be useful in performance analysis in production. Our approach inherits the advantages of texture based optimizations like textures-sets, mip-maps and level of detail which can be easily incorporated. Although we demonstrate on a fixed texture resolution, it can be tailored accordingly depending on the hardware constraints and rendering quality needed. This is in contrast to vertex based methods that provide vertex count as the only control knob and little control over level of detail.


A limitation of transfer textures is that inter-reflections essentially bake the lighting and BRDF i.e. they cannot be changed without re-computation. We note that the work of \cite{prt_sloan} also bakes BRDF (including albedo) into their transfer matrices for inter-reflections. We would like to address this issue for future extensions of this work.
\bibliographystyle{ACM-Reference-Format}
\bibliography{ms}

\end{document}